\documentclass[12pt]{article}

\usepackage[centertags]{amsmath}
\usepackage{amsfonts}
\usepackage{amssymb}
\usepackage{amsthm}
\usepackage{newlfont}
\usepackage{epsfig}
\usepackage{amscd}

\newcommand{\NN}{{\mathbb N}}

\newcommand{\RR}{{\mathbb R}}

\newcommand{\beq}{\begin{equation}}
\newcommand{\eeq}{\end{equation}}
\newcommand{\ba}{\begin{array}}
\newcommand{\ea}{\end{array}}
\newcommand{\bea}{\begin{eqnarray}}
\newcommand{\eea}{\end{eqnarray}}

\begin{document}

\begin{center}
{\large \sc \bf A hierarchy of integrable PDEs in 2+1 dimensions} 

\vskip 5pt
{\large \sc \bf associated with  $2$ - dimensional vector fields} 

\vskip 15pt

{\large  S. V. Manakov$^{1,\S}$ and P. M. Santini$^{2,\S}$}

\vskip 8pt

{\it 
$^1$ Landau Institute for Theoretical Physics, Moscow, Russia

\smallskip

$^2$ Dipartimento di Fisica, Universit\`a di Roma "La Sapienza" and\\
Istituto Nazionale di Fisica Nucleare, Sezione di Roma 1\\
Piazz.le Aldo Moro 2, I-00185 Roma, Italy}

\vskip 5pt

$^{\S}$e-mail:  {\tt manakov@itp.ac.ru, paolo.santini@roma1.infn.it}

\vskip 5pt

{\today}

\end{center}

\begin{abstract}
We study a hierarchy 
of integrable partial differential equations in $2+1$ dimensions 
arising from the commutation of $2$ - dimensional vector fields, and we construct 
the formal solution of the associated Cauchy problems using the Inverse Scattering 
Transform for one-parameter families of vector fields recently introduced in 
\cite{MS1,MS2}. Due to the ring property of the space of eigenfunctions, 
the inverse problem can be formulated in three distinguished ways; in particular, 
one formulation corresponds to a linear integral equation for a Jost eigenfunction, 
and another formulation is a scalar 
nonlinear Riemann problem for suitable analytic eigenfunctions. 
\end{abstract}

\section{Introduction}

In this paper we study a hierarchy 
of integrable partial differential equations (PDEs) in $2+1$ dimensions 
arising from the commutation of $2$ - dimensional vector fields, and we construct 
the formal solution of the associated Cauchy problems using the Inverse Scattering 
Transform (IST) for vector fields recently introduced in \cite{MS1,MS2}. 

The first nontrivial member of such hierarchy is the equation       
\beq
\label{MS}
\ba{l}
v_{xt}+v_{yy}=v_yv_{xx}-v_xv_{xy},~~~~~~~ 
v=v(x,y,t)\in\RR,~~~~~x,y,t\in\RR,
\ea
\eeq
introduced in \cite{Pavlov} and studied in \cite{Pavlov,Ferapontov,Duna} from different points of view, 
which admits the operator representation 
\beq
\label{operator_repr}
[\hat L,\hat M]=0
\eeq
in terms of the 2-dimensional vector fields \cite{Duna}
\beq
\label{L1L2-MS}
\ba{l}
\hat L\equiv \partial_y+(p+v_x)\partial_x, \\
\hat M\equiv \partial_t+(p^2+pv_x-v_y)\partial_x,
\ea
\eeq
where $p$ is a constant parameter.

Equation (\ref{MS}) is the $u=0$ reduction of the recently introduced \cite{MS1} integrable 
system:
\beq
\label{dKP-system}
\ba{l}
u_{xt}+u_{yy}=-(uu_x)_x-v_xu_{xy}+v_yu_{xx}, ~~~~u,v\in\RR,~~x,y,t\in\RR,\\
v_{xt}+v_{yy}=-uv_{xx}-v_xv_{xy}+v_yv_{xx}
\ea
\eeq
admitting the operator representation $[\tilde L,\tilde M]=0$ in terms  
of the following 3-dimensional vector fields 
\beq
\label{L1L2-syst}
\ba{l}
\tilde L\equiv \partial_y+(p+v_x)\partial_x-u_x\partial_p, \\
\tilde M\equiv \partial_t+(p^2+pv_x+u-v_y)\partial_x+(-pu_x+u_y)\partial_p.
\ea
\eeq
We remark that the $v=0$ reduction of the system (\ref{dKP-system}) 
is the celebrated dispersionless Kadomtsev-Petviashvili (dKP)      
(or Khokhlov-Zabolotskaya) equation:
\beq
\label{dKP1}
u_{tx}+u_{yy}+(uu_x)_x=0,~~~
u=u(x,y,t)\in\RR,~~~~~x,y,t\in\RR.
\eeq
arising in various problems of Mathematical Physics, and   
intensively studied in the recent literature (see, f.i., 
\cite{Zak,Kri,TT,DMT,K-MA-R,G-M-MA}). 

The inverse scattering problem for one - parameter families of multidimensional 
vector fields has been recently introduced in \cite{MS1}, and used to construct 
the formal solution of the Cauchy problem for a wide class of nonlinear PDEs in 
arbitrary dimensions \cite{MS1}, including the heavenly equation of Plebanski \cite{Pleb}, 
and for the system (\ref{dKP-system}) \cite{MS2}. The Hamiltonian constraint on 
the associated spectral data leading to the heavenly equation and to the dKP reduction of 
(\ref{dKP-system}) was also found in \cite{MS1} and \cite{MS2}. We remark that, in the case of 
Hamiltonian vector fields, an  
elegant and alternative integration scheme was already known in the literature \cite{Kri}. 

We also remark that, since the hierarchy of PDEs investigated in this paper arises from the commutation of 
$2$ - dimensional vector fields, according to the theory developed in \cite{MS1,MS2}, the associated 
inverse problem can be formulated in terms of a {\it scalar nonlinear Riemann problem}.

\section{Hierarchy of PDEs associated with $\hat L$}

Equation (\ref{MS}) is the first nontrivial member (corresponding to $n=2$) of the following 
 hierarchy of PDEs in $2+1$ dimensions 
arising from the commutation of $2$ - dimensional vector fields:
\beq
\label{hierarchy}
\ba{l}
v_{t_n}+\hat{\cal Q}^nv_x=0,~~~~~n\in\NN,
\ea
\eeq
where $\hat{\cal Q}$ is the recursion operator
\beq
\hat{\cal Q}\equiv \partial^{-1}_x\left( v_{xx}-\partial_y -v_x\partial_x \right)=
v_x-\partial^{-1}_x\partial_y -2\partial^{-1}_xv_x\partial_x.
\eeq
The first few flows are:
\beq
\ba{l}
v_{t_0}+v_x=0, \\
v_{t_1}-v_y=0, \\
v_{xt_2}+v_{yy}-v_yv_{xx}+v_xv_{xy}=0, \\
v_{xxt_3}-\left(v_y+v^2_x\right)_{yy}+\left[v_{xy}\left(v_y-v^2_x\right)+v_{xx}\partial^{-1}_x
\left(v_y+v^2_x\right)_{y}\right]_x=0.
\ea
\eeq
This hierarchy admits the operator representation
\beq
\label{operator_repr_n}
\ba{l}
\left[ \hat L,\hat M_n\right]=0,~~n\in\NN,
\ea
\eeq
where $\hat L$ is defined in (\ref{L1L2-MS}) and the 2-dimensional vector fields $\hat M_n$ are defined by
\beq
\label{M_n}
\ba{l}
\hat M_n\equiv\partial_{t_n}+\left( p^n+\sum\limits_{k=0}^{n-1}p^kA^{(n)}_{n-k}\right)\partial_x,
\ea
\eeq
with: 
\beq
\label{A_n}
\ba{l}
A^{(n)}_{k+1}=\hat{\cal Q}A^{(n)}_{k}, ~~1\le k\le n-1,~~~~A^{(n)}_{1}=v_x.
\ea
\eeq
To show it, we replace (\ref{L1L2-MS}a) and (\ref{M_n}) into (\ref{operator_repr_n}), obtaining the 
following polynomial (in $p$) equation:
\beq
v_{xt}+v_{xx}\left( p^n+\sum\limits_{k=0}^{n-1}p^kA^{(n)}_{n-k}\right)=
\sum\limits_{k=0}^{n-1}p^k(A^{(n)}_{n-k})_y+(p+v_x)\sum\limits_{k=0}^{n-1}p^k(A^{(n)}_{n-k})_x,
\eeq 
which must be satisfied $\forall p$. Equating to zero all powers of $p$, we obtain equations (\ref{A_n}) 
and (\ref{hierarchy}). 

Since $\hat{\cal Q}$ does not depend explicitly on $x$, its Lie derivative along $v_x$, the first 
vector field of the hierarchy, is zero; in addition, it is also possible to verify that $\hat{\cal Q}$ is a 
Nijenhuis (hereditary) operator. Therefore all the flows (\ref{hierarchy}) commute with each 
other \cite{FF,Magri}. 

We remark that the algebraic theory for the $2$ - dimensional recursion operator $\hat{\cal Q}$ 
is conceptually more similar to the theory for the $1$ - dimensional recursion operators 
associated with systems like the Korteweg - de Vries and nonlinear Shr\"odinger equations \cite{FF}, 
rather than to the theory for the $2$ - dimensional 
recursion operators associated with systems like the Kadomtsev - Petviashvili and Davey Stewartson 
equations \cite{SF}.  
No results are known though, at the moment, on the Hamiltonian and bi-Hamiltonian \cite{Magri} character 
of the hierarchy (\ref{hierarchy}).

\section{Inverse scattering transform}

Now we consider the Cauchy problem for the hierarchy (\ref{hierarchy}) and, in particular, 
for equation (\ref{MS}),     
within the class of rapidly decreasing real potentials $v$:
\beq
\label{localization}
v(x,y,t_n)\to~0,~~(x^2+y^2)\to\infty, ~~v\in\RR,~~~(x,y)\in\RR^2,~~t_n>0,
\eeq
interpreting $t_n$ as time and the other two variables $x,y$ as space variables.  
To solve such a Cauchy problem by the IST method developed in \cite{MS1,MS2}, 
we construct the IST for the operator $\hat L$, within the class of rapidly 
decreasing real potentials $v$, interpreting the operators $\hat M_n$ as  
time operators. 

The operator representation (\ref{operator_repr_n}) implies the existence of common 
eigenfunctions $f(x,y,t,p)$ of $\hat L,\hat M_n$ (the Lax pair): 
\beq
\label{Lax}
\hat L f=0,~~\hat M_n f=0,~~~n\in\NN.
\eeq
Since the Lax pair (\ref{Lax}) is made of vector fields,  
{\it the space of eigenfunctions is a ring}:  if $f_1,~f_2$ are two solutions 
of the Lax pair 
(\ref{Lax}), then an arbitrary differentiable function $F(f_1,f_2)$ of them 
is a solution of 
(\ref{Lax}).   

\subsection{Direct problem}

The localization (\ref{localization}) of the 
potential $v$ implies that, if $f$ is a solution of $\hat L f=0$, then 
\beq
\label{asymptf}
\ba{l}
f(x,y,p)\to f_{\pm}(\xi,p),\;\;y\to\pm\infty, \\
\xi:=x-py;
\ea
\eeq
i.e., asymptotically, $f$ is an arbitrary function of $\xi=x-py$ and $p$.

A central role in the theory is played by the real Jost eigenfunction $\varphi(x,y,p)$, 
the solution of $\hat L\varphi=0$ uniquely defined by the asymptotics 
\beq
\label{asympt-varphi}
\varphi(x,y,p)\to \xi,\;\;\;\;y\to -\infty,
\eeq
and equivalently characterized by the linear integral equation 
$\varphi=\xi+\hat G(-v_x\varphi_x)$, for the Green's function 
$G(x,y,p)=\theta(y)\delta(x-py)$.

The $y=+\infty$ limit of $\varphi$ defines the natural scattering datum $\sigma$ 
for $\hat L$:
\beq
\label{def-S}
\displaystyle\lim_{y\to +\infty}\varphi(x,y,p) \equiv 
{\cal S}(\xi,p)=\xi+\sigma(\xi,p).
\eeq

The direct problem is the transformation from the real potential $v$, function of the 
two real variables $(x,y)$, to the real scattering datum $\sigma$, function of the two real 
variables $(\xi,p)$. Therefore the mapping is consistent. In the small field limit, 
this mapping reduces to the {\it Radon transform} \cite{Radon}:
\beq
\sigma(\xi,p)=-\int_{\RR}v_{\xi}(\xi+py,y)dy.
\eeq
 
A crucial role in the IST theory for the vector field $\hat L$  
is also played by the analytic eigenfunctions $\psi_{\pm}(x,y,p)$, the solution of 
$\hat L\psi_{\pm}=0$  satisfying the integral equations 
\beq
\label{def-psi}
\ba{l}
\psi_{\pm}(x,y,p)=-
\int_{\RR^2}dx'dy'G_{\pm}(x-x',y-y',p)v_{x'}(x',y'){{\psi}_{\pm}}_{x'}(x',y',p)+\xi,
\ea
\eeq
where $G_{\pm}$ are the analytic Green's functions
\beq
\label{Green_analytic}
G_{\pm}(x,y,p)=\pm\frac{1}{2\pi i[x-(p \pm i\epsilon) y]}.
\eeq
The analyticity properties of $G_{\pm}(x,y,p)$ in the complex $p$ - plane 
imply that $\psi_{+}(x,y,p)$ and $\psi_{-}(x,y,p)$ are 
analytic, respectively, in the upper and lower halves of the $p$ - plane, with 
the following asymptotics, for large $p$:
\beq
\label{asympt-psi}
\psi_{\pm}(x,y,p)=\xi-\frac{v(x,y)}{p}+O\left(\frac{1}{p^2}\right),~~|p|>>1.
\eeq

It is important to remark that the analytic Green's functions (\ref{Green_analytic}) 
exhibit the following asymptotics for $y\to\pm\infty$:
\beq
\ba{l}
G_{\pm}(x-x',y-y',p)\to\pm\frac{1}{2\pi i[\xi-\xi'\mp i\epsilon]},\;\;y\to +\infty, \\
G_{\pm}(x-x',y-y',p)\to\pm\frac{1}{2\pi i[\xi-\xi'\pm i\epsilon]},\;\;y\to -\infty,
\ea
\eeq
entailing that {\it the $y=+\infty$ asymptotics of $\psi_{+}$ and $\psi_{-}$ are 
analytic 
respectively in the lower and upper halves of the complex plane $\xi$, while the 
$y=-\infty$ 
asymptotics of $\psi_{+}$ and $\psi_{-}$ are analytic respectively in the upper 
and lower 
halves of the complex plane $\xi$} (similar features have been observed first in \cite{MZ} and 
later in \cite{MS1,MS2}).

The Jost eigenfunction $\varphi$ and the constant eigenfunction $p$ form a basis 
in the ring of solutions of $\hat Lf=0$; thus any 
solution $f$ of $\hat Lf=0$ is a function of 
$\varphi$ and $p$. The analytic eigenfunctions $\psi_{\pm}$, in particular, possess the representations:
\beq
\label{varphi-psi}
\psi_{\pm}={\cal K}_{\pm}(\varphi,p)=\varphi+\chi_{\pm}(\varphi,p),
\eeq
defining the spectral data $\chi_{\pm}$. 

Since the $y\to -\infty$ limit of (\ref{varphi-psi}) reads:
\beq
\label{lim-varphi-psi}
\displaystyle\lim_{y\to -\infty}\psi_{\pm}-\xi=\chi_{\pm}(\xi,p),
\eeq
the above analyticity properties of the LHS of (\ref{lim-varphi-psi}) in the complex 
$\xi$ - plane imply that $\chi_{+}(\xi)$ and $\chi_{-}(\xi)$ are analytic respectively 
in the upper and lower halves  of the complex plane $\xi$, decaying at $\xi\sim\infty$ 
like $O(\xi^{-1})$. Therefore their Fourier transforms 
$\tilde{\chi}_{+}(\omega)$ and $\tilde{\chi}_{-}(\omega)$ have support respectively on 
the positive and negative $\omega$ semi-axes. 

The spectral data $\chi_{\pm}$ can be constructed from the scattering datum $\sigma$  
through the following linear integral equations 
\beq
\label{Fourier-varphi-psi}
\ba{l}
\tilde{\chi}_+(\omega,p)+\theta(\omega)\left(\tilde{\sigma}(\omega,p)+
\int_{\RR}d\eta ~\tilde{\chi}_+(\eta,p)Q(\eta,\omega,p)\right)=0,  \\
\tilde{\chi}_-(\omega,p)+\theta(-\omega)\left(\tilde{\sigma}(\omega,p)+
\int_{\RR}d\eta ~\tilde{\chi}_-(\eta,p)Q(\eta,\omega,p)\right)=0,
\ea
\eeq 
involving the Fourier transforms $\tilde{\sigma}$ and $\tilde{\chi}_{\pm}$ of 
$\sigma$ and ${\chi}_{\pm}$:
\beq
\label{Fourier-sigma}
\tilde{\sigma}(\omega,p)=\int_{\RR}d\xi\sigma(\xi,p)e^{-i\omega\xi},~~~
\tilde{\chi}_{\pm}(\omega,p)=\int_{\RR}d\xi{\chi}_{\pm}(\xi,p)e^{-i\omega\xi}
\eeq
and the kernel:
\beq
\label{def-Q}
Q(\eta,\omega,p)=\int_{\RR}\frac{d\xi}{2\pi}e^{i(\eta-\omega)\xi}
[e^{i\eta\sigma(\xi,p)}-1].
\eeq 
To prove this result, one first evaluates (\ref{varphi-psi}) at $y=+\infty$, obtaining 
\beq
\label{+lim-varphi-psi}
\left(\displaystyle\lim_{y\to \infty}\psi_{\pm}-\xi\right)=\sigma(\xi,p)+
\chi_{\pm}(\xi+\sigma(\xi,p),p).
\eeq 
Applying the integral operator $\int_{\RR}d\xi e^{-i\omega\xi}\cdot$ 
for $\omega>0$ and $\omega<0$ respectively to equations (\ref{+lim-varphi-psi})$_{+}$ and 
(\ref{+lim-varphi-psi})$_{-}$, using the above analyticity properties  
and the Fourier representations of ${\chi}_{\pm}$ and $\sigma$, one obtains equations 
(\ref{Fourier-varphi-psi}).

The reality of the potential: $v\in\RR$ implies that, for $p \in\RR$,  
$\overline{\varphi}=\varphi$, $\overline{\psi}_+=\psi_-$; consequently: 
$\overline{\sigma}=\sigma$, $\overline{\chi}_+=\chi_-$.

\subsection{Inverse problem(s)}

As in \cite{MS1,MS2}, due the ring property of the space of eigenfunctions, it is possible to 
construct three distinguished versions of the inverse problem, all based on equations 
(\ref{varphi-psi}). 

The first version is obtained subtracting $\xi$ from equations (\ref{varphi-psi})$_{-}$ and 
(\ref{varphi-psi})$_{+}$,  
applying respectively the analyticity projectors $\hat P_{+}$ and $\hat P_{-}$: 
\beq
\hat P_{\pm}\equiv \pm\frac{1}{2\pi i}\int_{\RR}\frac{dp'}{p'-(p\pm i\epsilon)}. 
\eeq
and adding up the resulting equations, to obtain the following 
nonlinear integral equation for the Jost eigenfunction $\varphi$:
\beq
\label{varphi-int-equ2}
\ba{l}
\varphi(x,y,p)+\frac{1}{2\pi i}\int_{\RR}\frac{dp'}{p'-(p+i\epsilon)}\chi_-\left(\varphi(x,y,p'),p'\right) - \\ 
\frac{1}{2\pi i}\int_{\RR}\frac{dp'}{p'-(p-i\epsilon)}\chi_+\left(\varphi(x,y,p'),p'\right)=x-py.
\ea
\eeq
Once $\varphi$ is reconstructed from (\ref{varphi-int-equ2}), given $\chi_{\pm}(\xi,p)$,  
the analytic eigenfunctions 
follow from (\ref{varphi-psi}), and the potential $v$ from equation (\ref{asympt-psi}). This inversion 
procedure was first introduced in \cite{Manakov1} and also used in 
\cite{MS1,MS2}.

The second version is the linear analogue of the nonlinear problem  
(\ref{varphi-int-equ2}), obtained {\it exponentiating the Jost and analytic eigenfunctions} used so far. 
Consider the following functions:
\beq
\Phi(x,y,p;\alpha)\equiv e^{i\alpha\varphi(x,y,p)},~~~
\Psi_{\pm}(x,y,p;\alpha)\equiv e^{i\alpha\psi_{\pm}(x,y,p)},~~~~\alpha\in\RR.
\eeq
Due to the ring property of the space of eigenfunctions, also $\Phi(x,y,p;\alpha)$ and 
$\Psi_{\pm}(x,y,p;\alpha)$ 
are eigenfunctions; $\Phi(x,y,p;\alpha)$ is a Jost eigenfunction characterized by the asymptotics 
$\Phi\to exp(i\alpha\xi),~y\to -\infty$, while $\Psi_{\pm}(x,y,p;\alpha)$ are analytic respectively 
in the 
upper and lower halves of the $p$ plane, with asymptotics: $\Psi_{\pm}=exp(i\alpha\xi)
[1-(\alpha v)/p+O(p^{-2})]$.

Exponentiating the representations (\ref{varphi-psi}), one obtains the {\it linear} 
expansions of the analytic 
eigenfunctions $\Psi_{\pm}$ in terms of the Jost eigenfunction $\Phi$:
\beq
\label{Phi-Psi}
\ba{l}
\Psi_{\pm}(x,y,p;\alpha)=\Phi(x,y,p;\alpha)+
\int_{\RR}d\beta K_{\pm}(\alpha,\beta,p)\Phi(x,y,p;\beta),
\ea
\eeq
where the new spectral data $K_{\pm}$ are defined in terms of $\chi_{\pm}$ by:
\beq
\label{K}
\ba{l}
K_{\pm}(\alpha,\beta,p)\equiv \int_{\RR}\frac{d\xi}{2\pi}e^{i(\alpha-\beta)\xi}
[e^{i\alpha\chi_{\pm}(\xi,p)}-1].
\ea
\eeq
Multiplying the equations (\ref{Phi-Psi})$_{+}$ and (\ref{Phi-Psi})$_{-}$ by 
$exp(-i\alpha\xi)$, subtracting $1$, applying respectively $\hat P_{-}$ and $\hat P_{+}$,  
and adding the resulting equations, one obtains the following {\it linear integral equation} for $\Phi$: 
\beq
\label{Phi-int-equ}
\ba{l}
\Phi(x,y,p;\alpha)+\frac{1}{2\pi i}\int_{\RR}\frac{dp'}{p'-(p+i\epsilon)}

\int_{\RR}d\beta K_-(\alpha,\beta,p')\Phi(x,y,p';\beta)e^{i\alpha(p'-p)y} - \\
~~ \\
-\frac{1}{2\pi i}\int_{\RR}\frac{dp'}{p'-(p-i\epsilon)}
\int_{\RR}d\beta K_+(\alpha,\beta,p')\Phi(x,y,p';\beta)e^{i\alpha(p'-p)y} =
e^{i\alpha(x-py)}. 
\ea
\eeq
Once $\Phi$ is reconstructed from (\ref{Phi-int-equ}) and, via (\ref{Phi-Psi}), $\Psi_{\pm}$ 
are also known, 
the potentials  are reconstructed in the usual way from the asymptotics of $\Psi_{\pm}$.

The reality constraint ($v=0$) implies, for $\alpha,p\in\RR$: $\overline{\Phi(\alpha)}=\Phi(-\alpha)$, 
$\overline{\Psi_+(\alpha)}=\Psi_-(-\alpha)$. Consequently: $\overline{K_+(\alpha,\beta)}=K_-(-\alpha,-\beta)$.

The third version of the inverse problem is a Riemann problem.  
Solving the algebraic system (\ref{varphi-psi})$_-$ with respect to $\varphi$: $\varphi={\cal L}(\psi_{-},p)$ 
(assuming local invertibility) and replacing   
this expression in the algebraic system (\ref{varphi-psi})$_+$, one obtains the representation of the 
analytic eigenfunction $\psi_{+}$ in terms of the analytic eigenfunction $\psi_{-}$:
\beq
\label{RH}
\psi_{+}={\cal R}(\psi_{-},p)=\psi_{-}+R(\psi_{-},p),~~p\in\RR,
\eeq
where ${\cal R}(\psi_-,p)={\cal K}_+({\cal L}(\psi_{-},p),p)$, 
which defines a {\it scalar nonlinear Riemann problem on the real $p$ axis}. The Riemann datum  
${\cal R}$ is therefore constructed from the data ${\cal K}_{\pm}$ by algebraic manipulation. Vice-versa, 
given $R$, one constructs the solutions $\psi_{\pm}$ of the nonlinear Riemann problem (\ref{RH}) and, 
via the asymptotics (\ref{asympt-psi}), the potential $v$.  

The reality constraint for $\cal R$ takes the form: 
${\cal R}(\overline{ {\cal R}(\bar{\xi},p)   },p)=\xi,~\forall\xi$, 
for $p \in\RR$. 

\vskip 5pt
\noindent
{\it Remark }. Dressing schemes can be formulated from the three different inverse problems 
presented in this paper in a straightforward way. 

\subsection{$t$ - evolution of the spectral data}

As $v$ evolves in time according to (\ref{MS}), the $t$-dependence of the  
spectral data $\sigma$, ${\chi}_{\pm}$, $R$ and $K_{\pm}$ is described by the following 
explicit formulas:
\beq
\label{t-dep-Sigma}
\ba{l}
\sigma(\xi,p,t)=\sigma(\xi-p^2t,p,0),~~~~~~{\cal S}(\xi,p,t)=\xi+\sigma(\xi-p^2t,p,0), \\
\chi_{\pm}(\xi,p,t)=\chi_{\pm}(\xi-p^2t,p,0),~~{\cal K}_{\pm}(\xi,p,t)=\xi+\chi_{\pm}(\xi-p^2t,p,0), \\
R(\xi,p,t)=R(\xi-p^2t,p,0),~~~~~~{\cal R}(\xi,p,t)=\xi+R(\xi-p^2t,p,0),  \\
K_{\pm}(\alpha,\beta,p,t)=K_{\pm}(\alpha,\beta,p,0)e^{i(\alpha-\beta)p^2t}.
\ea
\eeq
To prove it, we first observe that
\beq
\label{def-phi1}
\ba{l}
\phi(x,y,t,p)\equiv \varphi(x,y,t,p)-p^2t
\ea
\eeq
is a    common Jost eigenfunction of $\hat L$ and $\hat M$. The $y=+\infty$ 
limit of equation $\hat M\phi=0$ 
yields ${\sigma}_t+p^2{\sigma}_{\xi}=0$,  whose solution is (\ref{t-dep-Sigma}a). Analogously, 
\beq
\ba{l}
{\pi_{\pm}}(x,y,t,p)\equiv {\psi_{\pm}}(x,y,t,p)-p^2t
\ea
\eeq
are common analytic eigenfunctions of $\hat L$ and $\hat M$; therefore 
\beq
{\pi_{\pm}}={{\check{\cal K}}_{\pm}}(\phi,p),~~~~~~\pi_+=\check{\cal R}(\pi_-,p),
\eeq
for some functions ${\check{\cal K}_{\pm}}$ and $\check{\cal R}$ depending on $x,y,t$ 
only through $\phi$ and $\pi_-$. Comparing at $t=0$ these equations 
with equations (\ref{varphi-psi}) and (\ref{RH}), one expresses ${\check{\cal K}_{\pm}}$ and $\check{\cal R}$   
in terms of ${{\cal K}_{\pm}}$ and $\cal R$, obtaining equations (\ref{t-dep-Sigma}b,c). 
The $t$-evolution of $K_{\pm}$ is obtained replacing (\ref{t-dep-Sigma}b) in (\ref{K}).  

Analoguosly, one can obtain the explicit $t$ - dependence of the spectral data 
as $v$ evolves according to the other equations of the hierarchy (\ref{hierarchy}).

\section{One-parameter families of commuting \\ dynamical systems}

It is well-known (see, f.i., \cite{CH}) that linear first order PDEs 
like (\ref{Lax}),(\ref{L1L2-MS}),(\ref{M_n}) are intimately related to systems of ordinary 
differential equations describing their 
characteristic curves. For instance, the dynamical systems associated with the one-parameter family 
of vector fields $\hat L,\hat M$ in (\ref{L1L2-MS}) are: 
\beq
\label{flow}
\ba{ll}
\hat L:   & 
\frac{dx}{dy}=p+v_x(x,y,t), \\
\hat M:  & \frac{dx}{dt}=p^2+pv_x(x,y,t)-v_y(x,y,t).
\ea
\eeq
Therefore {\it equation (\ref{MS}) characterizes the class of functions $v$ for which the two first  
order dynamical systems (\ref{flow}) commute $\forall p$}.  
 
There is also a deep connection between the above IST and the time ($y$) - scattering theory for the 
commuting flows (\ref{flow}). Let $\phi(x,y,t,p)$ be the common eigenfunc tion of $\hat L$ and 
$\hat M$ defined in (\ref{def-phi1}); then, solving the system 
$\omega=\phi(x,y,t,p)$ with respect to $x$ (assuming local invertibility), one obtains the following 
common solution of (\ref{flow}):
\beq
\omega=\varphi(x,y,t;p)-p^2t~~\Leftrightarrow~~
x=r(\omega,y,t;p)~\sim~py+p^2 t+\omega,~~y\sim -\infty.
\eeq
The $y=+\infty$ limit of the solution $r(\omega,y,t;p)$:
\beq
x~\sim~py+p^2 t+\Omega(\omega,p),~~y\sim +\infty
\eeq
defines the time ($y$) - scattering datum $\Delta(\omega,p)=\Omega(\omega,p)-\omega$ 
of (\ref{flow}a), which is connected with the IST datum ${\cal S}$ by inverting the 
system $\omega={\cal S}(x-py-p^2t,p,0)$ with respect to $x$:
\beq
\omega={\cal S}(x-py-p^2t,p,0)~\Leftrightarrow~x-py-p^2t=\Omega(\omega,p).
\eeq
As a byproduct of the IST of 
this paper one can reconstruct, from the scattering 
datum $\Delta(\omega,p)$ of the one-parameter family of dynamical systems (\ref{flow}a), 
the function $v$.

Similar considerations can be made for the dynamical systems associated with the whole 
hierarchy of one-parameter family of operators $\hat L,\hat M_n,~n\in\NN$.

\vskip 10pt 
\noindent
{\bf Acknowledgments}. P. M. Santini was supported by the INFN grant 2006, and S. V. Manakov by the RFBR 
grants 04-01-00508, 06-01-90840, and 06-01-92053. PMS thanks M. Dunajski for pointing out few references 
in which equation (\ref{MS}) was studied.

\end{document}